\documentclass[twocolumn,
pra,
 showpacs,floats,preprintnumbers,amsmath,amssymb]{revtex4}
\usepackage{epsfig}
\usepackage{amsmath}
\renewcommand{\(}{\left(}
\renewcommand{\)}{\right )}

\renewcommand{\]}{\right ]}

\def\pa{\partial}

\def\bea{\arraycolsep .1em \begin{eqnarray}}
\def\eea{\end{eqnarray}}

\def\vp{{\bf p}}

\def\vk{{\bf k}}

\let\om=\omega
\let\be=\beta
\let\no=\nonumber

\def\eq#1{Eq.(\ref{#1})}
\def\refr#1{\cite{#1}}

\def\s0#1#2{\mbox{\small{$ \frac{#1}{#2} $}}}
\def\0#1#2{\frac{#1}{#2}}

\def\anp#1#2#3{Adv.\ Nucl.\ Phys. \ {\bf #1}, #2 (#3)}
\def\cpl#1#2#3{Chin. Phys. Lett. {\bf #1}, #2 (#3)}
\def\ctp#1#2#3{Commun. Theor. Phys. {\bf #1}, #2 (#3)}
\def\plb#1#2#3{Phys. Lett. {\bf B #1}, #2 (#3)}

\def\pra#1#2#3{Phys. Rev.  {\bf A #1}, #2 (#3)}

\def\prc#1#2#3{Phys. Rev.  {\bf C #1}, #2 (#3)}

\def\prl#1#2#3{Phys. Rev. Lett. {\bf #1}, #2 (#3)}
\def\ann#1#2#3{Ann. Phys. (N.Y.) {\bf #1}, #2 (#3)}
\def\anp#1#2#3{Adv. Nucl. Phys. {\bf #1}, #2 (#3)}

\def\epl#1#2#3{Europhys.\ Lett.{\bf #1}, #2 (#3)}
\def\prep#1#2#3{Phys.\ Rep.\ {\bf #1}, #2 (#3)}

\def\ijmpb#1#2#3{Int.\ J.\ Mod.\ Phys.\ {\bf B #1}, #2 (#3)}

\def\ijmpe#1#2#3{Int.\ J.\ Mod.\ Phys.\ {\bf E #1}, #2 (#3)}

\def\rmp#1#2#3{Rev.\ Mod.\ Phys.\ {\bf #1}, #2 (#3)}

\def\sci#1#2#3{Science\ {\bf #1}, #2 (#3)}

\begin{document}
\title{Quantum rearrangement and self-consistent BCS-BEC crossover thermodynamics}
\author{Ji-sheng Chen$^a$\footnote{chenjs@iopp.ccnu.edu.cn}  ~~~~~
Chuan-ming Cheng$^b$~~~~~~~Jia-rong Li$^a$~~~~~~~Yan-ping Wang$^a$}
\affiliation{$^a$ Physics Department $\&$ Institute of Particle
Physics, Central China Normal University, Wuhan 430079, People's
Republic of China} \affiliation{$^b$ Physics Department, Yun Yang
Teachers College, Shi Yan 442700, People's Republic of China}
\begin{abstract}
Based on previous works,
    analytical calculational procedures for dealing with the
    strongly interacting fermions ground state are further developed
    through a medium dependent potential in terms of the Bethe-Peierls contact interaction model.
The methods are exact in the unitary limit regime and they lead
    to the self-consistent equations analogous to the Hartree ones.
The single particle energy spectrum rearrangement effects on the
    thermodynamics due to the Hugenholtz-van Hove theorem constraint are addressed.
These effects lead to an additional instantaneous correlation
    potential contribution to the system physical chemical potential
    and pressure,
    i.e., equation of state,
    which enforces the classical thermodynamic consistency.
The Dyson-Schwinger equations represent implicitly the various
Bethe-Goldstone expansion ones. In a thermodynamically
self-consistent way,
    the universal dimensionless factor is analytically calculated to be $\xi=\049$,
    which defines the ratio of
    the unitary fermions energy density to that of the ideal
    non-interacting ones at $T=0$.

\end{abstract}


\maketitle
\section{Introduction}
The strongly interacting matter physics serves currently as a
    pivot topic in the fundamental many-body theory or statistical mechanics context.
An interesting observable that characterizes this kind of physics is
    the unitary limit thermodynamics.

Taking the Feshbach resonance techniques,
    it is now possible to magnetically
    tune the two-body inter-particle interaction strength.
Increasing the scattering length of fermions from $-\infty$ to
    $+\infty$ resulting in bound boson systems can exhibit the
    crossover from Bardeen-Cooper-Schrieffer(BCS) to Bose-Einstein condensation(BEC).
This will lead to the universal thermodynamics of
    the Bethe-Peierls contact interaction fermions system with a zero-range,
    infinite scattering length
    interaction\refr{Bertsch1999,Baker2001,Heiselberg2000,Ho2004,Schwenk2005,Xiong2005,Horowitz2006,physics/0303094,Castin2006,Stringari2007,Bloch2007}.
To solve the universal resonant superfluidity thermodynamic problem,
    at least in principle, more theoretical efforts are urgent to
    understand the detailed dynamics by going beyond the mean field theory\refr{Pethick}.

At the crossover point, the dimensionless coefficient $\xi$
    relates the total energy $E/N =\xi \035 \epsilon_f$ with the Fermi kinetic energy
    $\epsilon_f=k_f^2/(2m)$.
Here, $m$ is the bare fermion mass while $k_f$ is the Fermi
momentum. The determination of the fundamental constant
    $\xi$ as a challenging many-body topic attracts much attention
    theoretically/experimentally\refr{Steele,physics/0303094,Astrakharchik2004,Bulgac2005,Hu2007}.
It is presently accepted to be in the region $0.40-0.46$. Compared
    with the old values,
    the updating experiments
    converge more towards $\xi\sim 0.42-0.46$\refr{Partridge2006,Stewart2006,Joseph2006}.

The ultra-cold fermion atoms gas physics offers a plausible
    perspective in looking for the general nonperturbative statistical field theory methods.
In the strongly interacting regime $k_f|a|>>1$, especially right at
    the unitary limit point $a=\pm \infty$,
    the infrared singular nature of the two-body scattering amplitude $f_0(k,a)=i/k$ (with $k$ being
    the relative wave-vector magnitude of the colliding particles) rules
    out any conventional perturbative expansion techniques.
It is generally believed that there is not a simple analytical
    method to compute the universal coefficient $\xi$, although this
    topic stimulates many growing arduous theoretical efforts.
By noting that an exact many-body theory concerns the behavior at
    unitarity is presently not available, it is not surprising that the
    existed analytical results about the universal coefficient $\xi$
    differ from each other considerably($\xi\sim 0.2 - 0.6$).
It is worthy noting that there are still uncertain discrepancies
    among the powerful quantum Monte-Carlo works themselves as well as
    the experimental ones existed in the
    literature\refr{physics/0303094,Astrakharchik2004,Bulgac2005}\refr{Partridge2006,Stewart2006,Joseph2006}.

In Refs.\refr{chen2006},
    we make an almost first principle
    detour to attack the universal constant $\xi$ with the general
    Dirac quantum many-body formalism.
The analytical circumvention
    is motivated by the strongly instantaneous Coulomb correlation
    thermodynamics discussions in a compact nuclear confinement
    environment\refr{chen2005,chen2005-1,chen2005-2} based on the relativistic continuum nuclear
    many-body framework\refr{walecka1974,Serot1986}.
The infrared correlation characteristic in the unitary limit
    promotes us to use the \textit{universal} strongly instantaneous long range
    interaction to attack the infinite contact interaction thermodynamics.
The obtained analytical result $\xi$ is comparable with other
    theoretical quantum Monte-Carlo attempts. It is derived through a medium dependent interaction
     by accounting for the interior correlation with a counteracting or anti-correlation external source static potential.

However, the auxiliary quantities pressure $\tilde{P}$ and
    chemical potential $\tilde{\mu}$ do not obey the profound
    universal hypothesis in the unitary limit with $T=0$\refr{Ho2004}.
According to the universal hypothesis at unitarity with $T=0$,
    the ratio of pressure to energy density should be still that of
    non-interacting fermion gas.
Therefore, the physical pressure should be $P =\xi P_{ideal}$
    instead of  $\tilde{P}=\016 P_{ideal}$ obtained in
    Refs.\refr{chen2006} and consequently affects the important sound speed.
The chemical potential should be $\mu =\xi \epsilon_f$ instead of
    $ \tilde{\mu}=\mu ^*+{2\pi
    a_{eff}}n/{m}=\epsilon_f-{\pi^2}n/{(mk_f)}=\013\epsilon_f$.
Intriguingly, the physical Helmholtz free energy density is not
    affected.
At $T=0$, it can be $ f=-\tilde{P}+\tilde{\mu }n =\xi \035
    n\epsilon_f$. Therefore the underlying energy(-momentum) conservation
    law is still exactly ensured even with the respective
    discrepancy of the auxiliary pressure and chemical potential.
This observation instructs us to reanalyze the physical pressure and
    system chemical potential carefully.

The Hugenholtz-van Hove(HvH) theorem characterizes the fundamental
    thermodynamic relation between energy density and
    pressure\refr{Hugenholtz}.
It is well known that any effective medium dependent interaction
    with Brueckner-Bethe-Goldstone(BBG) techniques can easily violate
    this relation.
The standard many-body technique to remove this tortuous difficulty
    is considering the quantum rearrangement effects of the single
    particle energy spectrum on the physical chemical potential and
    pressure\refr{Brown2002,Baker1971,Rios2005,Nakada2006}.
The role of the single particle energy spectrum shift will be
    explored in the strongly interacting unitary limit.

To elucidate the physical chemical potential and pressure(equation
    of state) at both finite temperature and density transparently with
    a more concise non-relativistic recipe for the ultra-cold atomic
    physics is the present main motivation.
Up to now, quantum statistical field theory of the strongly
    interacting matter at finite temperature and density is still a
    challenge.
More concretely, the exact determination of equation of state plays
    an important role in looking for the detailed knowledge about the
    collective excitation modes measured with the ultra-cold atomic physics experiments.

The present work is organized as follows. In Sec. \ref{sec2},
    the adopted low energy effective theory formalism and corresponding nonperturbative approach for the
    strongly interacting unitary gas thermodynamics are described.
Through taking into account the single particle energy rearrangement
effect, the unitary fermions thermodynamics with the dynamically
    self-consistent Dyson-Schwinger equations is presented in Sec. \ref{sec3}.
Moreover, the universal thermodynamics at unitarity with $T=0$ as
    well as the concrete
    comparisons with existed results
    is discussed in Sec. \ref{sec4}.
Section \ref{sec5} includes the discussion and prospective remarks.
The summary conclusion is made in the final Sec. \ref{sec6}.

In this work,
the calculations are performed in the non-relativistic
    contact interaction formalism.
Throughout the paper, the natural units $\hbar =k_B=1$ are employed.

\section{Formalism}\label{sec2}
\subsection{In-medium effective action}

The usual bare Bethe-Peierls Hamiltonian with a global $U(1)$
    symmetry addressing the many-body ground state energy can be found in the literature.
The in-medium behavior associated with the many-body correlation
    characteristic is the key,
    while a nonperturbative approach is crucial for the strongly interacting unitary regime physics.
Motivated by previous works,
    we work with the medium-scaling
    universal functional version
\bea\label{Hamiltonian} H=&&-\int d^3x
    \psi_\alpha ^*(x)
    (\0{\nabla ^2}{2m}-\mu_{r\alpha}[n])\psi _\alpha (x)\no\\
    &&+\0{U_{eff}^*[n]}2 \int d^3x\psi_\alpha ^*(x)\phi^*_\beta (x)
    \psi_\beta(x)\psi_\alpha (x). \eea

In \eq{Hamiltonian},
    m is the bare fermion mass with $\alpha,\be=\uparrow, \downarrow$
    representing the (hyperfine-)spin projection.
The physical many-body renormalized interaction potential
    $U_{eff}^*[n]$ is a functional of density and temperature,
    from which one can define the in-medium
    $S$-wave scattering length $a_{eff}$
\bea U_{eff}^*[n,T]\equiv \0{4\pi a_{eff}[n,T]}{m}. \eea The bare
interaction matrix $U_{eff}^*[n,T]\rightarrow U_0$ is directly
related with the free-space scattering length $a$ with the low
energy Born approximation for the zero range contact interaction.

Considering the medium influence of the surrounding environment on
    the single particle energy spectrum properties, we have included
    the rearrangement term $\propto\mu_{r\alpha}[n]$ in \eq{Hamiltonian}.
It is introduced through the
    auxiliary quantity ${\tilde \mu_\alpha}$ indicated by the tilde symbol
    $\tilde{\mu}_\alpha[n]=\mu_\alpha -\mu_{r\alpha}[n]$.
It means that the collective correlation effects are further
    accounted for as a single particle potential formalism in
    the spirit of density functional theory\refr{Kohn1965,Brown2002},
    while the true Lagrange multiplier is $\mu_\alpha$.
As a functional of density and temperature,
    its concrete medium
    dependence can be derived analytically from the thermodynamic
    relation for the fixed particle number density.

\subsection{Effective potential determined by the counteracting environment}

Theoretically, the derivation of the many-body renormalized
    effective interaction $U_{eff}^*$ is beyond the bare Bethe-Peierls
    contact interaction Hamiltonian itself,
    i.e., it should be derived
    from the more underlying physical law\refr{Brown2002}.

Firstly, from the point of view of the contact interaction fermions
physics,
    it is extremely instructive to recall the primary intermediate
    vector boson (IVB) hypothesis in weak interaction theory.
It is well known that the local intermediate vector boson  theory is
    related with the current-current(CC) contact interaction version
    through the corresponding connection: ${g^2_W}/{m_W^2}\equiv
    \0G{\sqrt{2}}$.
In the low energy limit $k\rightarrow 0$,
    the two IVB and CC theories are ingrainedly identical.
To model the low energy unitary limit physics, we ``let"
    ${g^2_W}/{m_W^2}\rightarrow \infty$ by taking $g_W$ to be an
    arbitrary \textit{large} charge or with an arbitrary \textit{small}
    mass gap $m_W$.
Consequently, the counteracting environment associated with an
    additional $U(1)$ conserved ``electric charge" can be introduced to
    attack the challenging infinity.
Constrained by the opposite charge neutralizing background, the
    effective interaction $U_{eff}^*$ is found to appear as an
    instantaneously anti-screening formalism.
This particular analogy and/or discursion makes the coupling
    constant $g_W$ enter simultaneously in the denominator and numerator
    of fractions in the relevant thermodynamic expressions\refr{chen2006}.
The assumed electric charge $g_W$ is reduced by the
    physical constraint or charge neutralizing stability condition and its
    concrete magnitude does not affect the final unitary physical results.
Therefore, there are not any regularization or expansion parameters
        violating the intriguing unitarity in gauging the infrared singular
        divergence.
This spirit will be further explored in this more concise
    non-relativistic recipe.

Moreover, corresponding to counteracting the deviation tendency of
    charge neutralizing,
there exists a similar general Le Chatelier's principle in
    thermodynamics.
This profound stability principle accounts for frustrating an
    instantaneous departure from equilibrium with an alternating
    \textit{minus} function of the
    surrounding environment\refr{Landau}.
First of all, as we will see, the counteracting-antiscreening
    formalism makes it possible for us to use the algebra equations to
    characterize the classical thermodynamics with the significant
    quantum correlation effects,
    but of crucial importance is that the many-body renormalization
    effects are introduced through a mirror background with an
    alternating minus function to avoid the theoretical double counting trouble.

\begin{figure}[ht]
        \centering
        \psfig{file=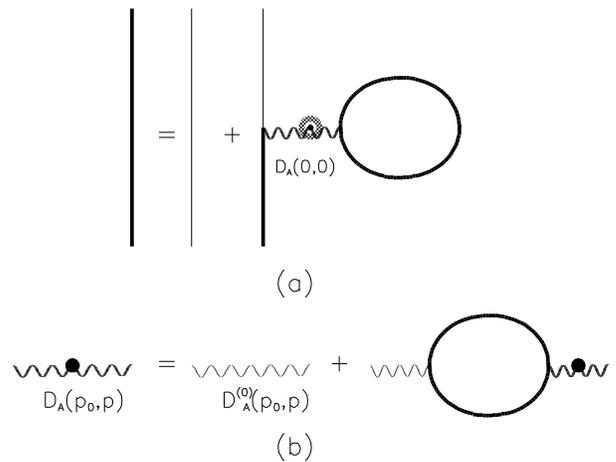,width=8.cm,angle=-0}
        \caption{
        \small
Generalized self-consistent Dyson-Schwinger equations incorporating
    the Landau pole contributions associated with the Fermi surface
    (anti-)correlation instability\refr{chen2006}:
(a), The full fermion propagator determined with
\textit{antiscreening}
    interaction potential in instantaneous approximation
    $U^*_{eff}[n]\equiv{U_0}/{(1+\Pi ^*(0,|{\bf p}|\rightarrow0)
    U_0)}$; (b), The boson propagating calculated with the full fermion
    propagator. }\label{fig1}
\end{figure}

The following calculations based on \eq{Hamiltonian} can be
    characterized by the comprehensive Feynman diagrams presentation with the
    generalized coupled Dyson-Schwinger equations(DSEs) formalism as indicated by Fig.\ref{fig1}.

The polarization tensor $\Pi ^*(\om_n, {\bf p})$ is calculated
    through the non-relativistic random phase approximation(RPA) with
    the full fermion propagator ${\cal G}(\om_n,{\bf k})$ as described
    by Fig.\ref{fig1}b
\bea \Pi^* (\om_n, {\bf p})&&=T \sum _{\om_1} \int _k {\cal G}(\om
    _1, {\bf k})
    {\cal G}(\om _1+\om _n, {\bf k}+{\bf p})\no\\
    &&=-\int _k \0{f_{{\bf k}+{\bf p}}-f_{\bf k }}{i \om _n -(\0{({\bf
    k}+{\bf p})^2}{2m}-\0{{\bf k}^2}{2m})},
\eea
with $\int _k=\int d^3{\bf k}/(2\pi)^3$.
Here, $f_{\bf k }$ is
the Fermi-Dirac distribution function given explicitly in
\eq{fermi-dirac}.

After employing the analytical continuation for the Matsubara
    frequency $\om _n$ according to $i\om _n\rightarrow \om +i\eta$, the
    $\Pi ^*(\om, {\bf p})$ determines the dispersion relation of the
    excitation mode A (the composite boson/dimer) in-medium propagating.
The polarization function obtained in this way is the standard
    in-medium Lindhard density-density or spin-spin response function for
    the one component fermions gas\refr{Pethick}.
In this work, we care about the low energy long wavelength
    thermodynamics of the fully pairing correlation occasion, which is
    related to the static infrared limit
\bea\label{debye} \Pi^* (0, |{\bf p}|\rightarrow
    0)=-\0{m_D^2(T,\lambda)}{2}=\01T(\0{m T}{2 \pi })^{3/2} \mbox{Li}_{1/2}[y].
\eea

The $m_D^2$ appearing in \eq{debye} is the physical non-relativistic
    Debye screening mass squared with unit electric charge and reflects
    the quantum fluctuation contributions consequently.
Its magnitude is the spin-spin or density-density correlation
    response susceptibility in the static infrared limit.
At $T\rightarrow 0$, $m_D^2$ can be reduced to the familiar
    formalism
\bea
    m_D^2|_{T=0}=\0{k_fm}{\pi ^2}.
\eea

Different from the multiplier-system physical chemical potential
$\mu$, the symbol $\lambda $ employed in above equations is the
    effective(total) chemical potential.
It characterizes the additional many-body collective correlation
    effects and will be defined explicitly in the following discussion
    for the full fermion propagator.
Here and afterwards, the function identifier $\mbox{Li}_i[y]$ stands
for
    the standard Fermi integral mathematica polylogarithm function
    $polyLog[i,y]$.
Meanwhile, we will use the shorting notation $y=-e^{\0{\lambda}T}$
    for the sake of conciseness throughout the paper.

With the polarization function $\Pi^* (0, |{\bf p}|\rightarrow 0)$,
    the rearrangement interaction matrix is derived to be
\bea \label{eff-p} U^*_{eff}[n,T]&&\equiv\0{U_0}{1+\Pi
^*(0,|\vp|\rightarrow0) U_0}=\0{U_0}{1-\0{m_D^2}2 U_0}. \eea At
$T=0$, it can be reduced to \bea
U^*_{eff}[n]=\0{U_0}{1-N(\epsilon_f) U_0}. \eea The
$N(\epsilon_f)=k_fm/(2\pi^2)$ is the un-perturbated density of
states on the Fermi surface for one component
fermions\refr{Pethick}, which is one half
    of the physical Debye screening mass squared.
Without a loss of generality and conformity, the screening mass
    squared notion $m_D^2$ instead of the density of states will be used
    to characterize the general finite temperature thermodynamics.

Compared with the conventional screened interaction customarily
    applied in the weak coupling perturbative loop diagram ring/ladder
    resummations
\bea\label{ind-p} U_{ind}(\om_n, {\bf p})&&=\0{U_0}{1-\Pi^*(\om_n,
    {\bf p}) U_0},\eea
there is a key dynamically mirrored \textit{minus} sign difference
    in the denominators of \eq{eff-p} and \eq{ind-p},
which leads to the quite different physical motivations and
calculational details.

\section{Thermodynamics with generalized Dyson-Schwinger equations}\label{sec3}
\subsection{Full fermion propagator and energy density}

The full fermion propagator ${\cal G}(\om_n,{\bf k})$ given with the
    instantaneously anti-screening interaction matrix tells us all the
    information of the ground state energy or quasi-particle
    distribution function.
It is calculated with the proper self-energy
    $\Sigma ^*$ through the instantaneous approximation as indicated by
    Fig.\ref{fig1}a
\bea\label{propagator}
    {\cal G}_\uparrow&&={\cal G}_{0\uparrow}+{\cal G}_{0\uparrow}\Sigma ^* {\cal G}_\uparrow,\\
    \Sigma ^*(0,0)&&=U_{eff}^*(0,|\vk|\rightarrow
    0){n_\downarrow}=U_{eff}^*[n]\0{n}2.
\eea Here, $n$ is the total particle number density
    $n=n_\uparrow+n_\downarrow=2n_\uparrow$ for the unpolarized/fully
    pairing symmetric trapped dilute atomic Fermi gas.

The self-energy $\Sigma ^*(0,0)$ is four-momentum independent in the
    instantaneous approximation.
It means that the self-energy will shift the chemical potential of
    the particle distribution function,
    i.e., the correlation
    contributions to thermodynamics are characterized through the phase
    space energy-momentum distribution \textit{deformation}.

From the full fermion propagator with \eq{Hamiltonian}, the
    particle number and energy densities are given by\refr{Walecka}
\bea
    \0NV&&=T\int _k\sum _n e^{i\om _n \eta }\mbox{tr} {\cal G}(\om_n,{\bf k}),\\
    \0EV&& =T\int _k\sum _n  e^{i\om _n \eta } \012 (i\om _n +\0{{\bf
    k}^2}{2m}+\tilde{\mu} )\mbox{tr} {\cal G}(\om_n,{\bf k}).
\eea
Although we once again consider the linear contact formalism,
    the lowest order diagrams characterize the
    thermodynamics with the highly nonlinear/turbulent
    contributions of the density and temperature fluctuations due to taking the medium dependent interaction
    potential\refr{Brown2002}.

After summing the Matsubara discrete frequency $\om _n$ with the
    residual theorem, the particle density $n=N/V$ and energy density
    $\epsilon=E/V$ are reduced to the compact
    formalisms
\begin{subequations}
    \begin{align}\label{number}
    n&=2\int _k f_{\bf k}=-2 (\0{m T}{2 \pi })^{3/2} \mbox{Li}_{3/2}[y],&\\
    \label{energy}\epsilon&=2\int_k \0{{\bf k}^2}{2m}f_{\bf k}+\0{\pi a
    _{eff}}{m}n (T,\lambda)^2&\no\\
    &=-3T(\0{m T}{2 \pi })^{3/2} \mbox{Li}_{5/2}[y]+\0{\pi a _{eff}}{m}n^2,&
    \end{align}
\end{subequations}
with the entropy density $s=S/V$
\bea\label{entropy}
    s&&=-2\int _k \[f_{\bf k}\ln f_{\bf k}+(1-f_{\bf k})\ln (1-f_{\bf k})\]\no\\
    &&=\01T(\0{m T}{2 \pi })^{3/2}\(2 \lambda
    \mbox{Li}_{3/2}[y]-5T\mbox{Li}_{5/2}[y]\).
\eea

The in-medium effective scattering length is \bea
    a_{eff}=\0a{1-\0{2\pi  m_D^2}{m}a}.
\eea
In deriving above equations, we have defined the quasi-particle
    Fermi-Dirac distribution function with
     the total/effective chemical
    potential $\lambda$ through Fig.\ref{fig1}a with \eq{propagator}
\bea\label{fermi-dirac} f_{\bf k}&&=\0{1}{e^{\beta
    (\0{{\bf k}^2}{2m}-\lambda)}+1},\\
    \lambda&&=\tilde{\mu}-\Sigma^*(0,0),
\eea
with $\beta =1/T$ being the inverse temperature.

The expressions for the entropy density \eq{entropy} and the first
    energy density term in \eq{energy} are similar to those of the
    non-interacting fermion gas.
The total chemical potential $\lambda$ is determined for the given
    particle number density $n$ and temperature $T$. The above algebra
    equations are completely \textit{closed}.

\subsection{Physical pressure and system chemical potential with
rearrangement}

From the Helmholtz free energy density $F/V=f=\epsilon -Ts$,
    the pressure is given by
\bea\label{pressure-d} P=-\(\0{\pa F}{\pa
    V}\)_{T,N}=-\(\0{\pa\0FN}{\pa \0VN}\)_T=n^2\(\0{\pa \0fn}{\pa n
    }\)_T.
\eea
At $T=0$ with entropy density $s=0$, $P$ is reduced to the
    important formalism\bea\label{p-e} P=n^2\0{\pa }{\pa n}
    \(\0{\epsilon}{n}\).
\eea
The equation \eq{p-e} determines the
    fundamental relation between the energy density and pressure
    according to the HvH theorem\refr{Hugenholtz},
    which plays a vital
    role for characterizing the symmetric nuclear matter saturation property.

As widely discussed in the nuclear many-body literature, the HvH
    theorem constraint will lead to the rearrangement terms in the
    pressure and chemical potential dealt with any kind of in-medium
    interaction potential\refr{Brown2002,Baker1971,Rios2005,Nakada2006}.
Presently, their exact analytical expressions can be derived in a
    thermodynamic way instead of doing the numerical
    interpolation\refr{chen2002}.

Let us further address the pressure $P$.
From \eq{pressure-d} given
with the medium dependent interaction potential,
one can have
\bea\label{pressure} P=&&\0{\pi a_{eff}}{m}n^2+{\cal C}(T,\lambda)
    \(\0{2\pi a_{eff}}{m}\)^2 n^3+P_{ideal}(T,\lambda),\no\\
\eea with
\bea
    P_{ideal}(T,\lambda)=&&-2T\(\0{mT}{2\pi}\)^{3/2}\mbox{Li}_{5/2}[y].
\eea
In \eq{pressure}, the first term represents the expected direct
    contribution of the induced interaction, while the last one results
    from the Fermi kinetic energy.
The essential second term is due to the rearrangement contributions,
    which can be easily neglected.
It is worthy noting that this rearrangement term is vanishing within
    the standard mean field theory approach\refr{chen2002}.

The discussion for the physical system chemical potential/multiplier
    is in line with that of pressure
\bea \mu =\(\0{\pa F }{\pa N }\)_{T,V}=\(\0{\pa \0{F}{V} }{\pa \0NV
    }\)_{T}=\(\0{\pa(\epsilon -Ts)}{\pa n}\)_T.
\eea
Its analytical expression can be further reduced to
\bea\label{chemical} \mu=\0{2\pi a_{eff}}{m}n+{\cal C}
(T,\lambda)\(\0{2\pi a_{eff}}{m}\)^2 n^2+\lambda. \eea
There is an
    additional(second) term in the system chemical potential resulting
    from the intriguing rearrangement effect of the single-particle degrees of freedom
    analogous to that of pressure.

Explicitly, the rearrangement terms in the pressure \eq{pressure}
    and chemical potential \eq{chemical} are similar.
For the general finite temperature occasion, the rearrangement
    factor ${\cal C} (T, \lambda)$ appearing in \eq{pressure} and \eq{chemical} is
    given by
\bea {\cal C} (T,\lambda)\equiv m_D\(\0{\pa m_D}{\pa
    n}\)_T=\0{\mbox{Li}_{-1/2}[y]}{2 T\mbox{Li}_{1/2}[y]}.
\eea

From the pressure and chemical potential analytical expressions, one
can see
    that the rearrangement effects play the positive ``repulsive" role.
These effects prevent the system from collapsing in the strongly
    coupling attractive unitary limit.
Furthermore, the in-medium
    spontaneously generated rearrangement effects are exactly
    counteracted(easily neglected) in the Helmholtz free energy density
    $f=\epsilon -T s=-p+\mu n $.
This reflects an important physical fact that the in-medium
    off-shell(spontaneous Lorentz violation) formalism can approach the
    quantum correlation effects on the universal thermodynamics,
    while the thermodynamics first law constraint ensures the energy
    conservation.
Exactly, the classical thermodynamic consistencies are satisfied
    with the generalized self-consistent DSEs.

\section{Universal thermodynamics in the unitary limit with $T=0$}\label{sec4}

From the general analytical expressions, one can give the $T=0$
    results with the in-medium interaction
\bea
    U_{eff}^*=\0{U_0}{1-\0{k_fm}{2\pi^2}U_0}\rightarrow
    a_{eff}=\0{a}{1-\0{2k_f a}{\pi}}.
\eea
With $n={k_f^3}/{(3\pi^2)}$ and the rearrangement factor
\bea
    {\cal C}(\epsilon_f)=m_D\0{\pa m_D}{\pa n}=\0{1}{4\epsilon_f},
\eea
    the physical internal energy density \eq{energy}, pressure
    \eq{pressure} and the system chemical potential \eq{chemical} are
\begin{subequations}
    \begin{align}
         \0E{N\epsilon_f}
        &=\013\0{2 a k_f}{\pi-{2 a k_f}}+\035,&\\
        \0{PV}{N\epsilon_f}&=\013\0{2 a k_f}{\pi-{2 a k_f}}+\019 \(\0{2 a k_f}{\pi-{2 a k_f}}\)^2+\025,&\\
         \0{\mu}{\epsilon_f}&=\023 \0{2 a k_f}{\pi-{2 a k_f}}+\019 \(\0{2 a k_f}{\pi-{2 a k_f}}\)^2+1.&
     \end{align}
\end{subequations}

In the strongly interacting universal unitary limit
    $|a|\rightarrow\infty$,
    the ratios of energy density, pressure and
    chemical potential to those of non-interacting fermion gas are
    $\xi=\049$.
This $\xi=\049$ is exactly consistent with the existed theoretical
    results\refr{physics/0303094,Steele,Bulgac2005} and in agreement
    with the updating experimental
    ones\refr{Partridge2006,Stewart2006,Joseph2006}.
In the
    weak coupling limit, the analytical thermodynamic expressions
    readily return to the lowest order mean field theory results which neglect the quantum fluctuation/correlaion contributions.

At unitarity and with $T=0$, the pressure is
\bea P=\04{45m}(3
    \pi^2)^{2/3}n^{1+2/3}=\xi P_{ideal},
\eea
with which the sound speed is calculated to be
\bea
    v=\sqrt{\0{\pa P}{m \pa n}}=\023 v_{FG}=\sqrt{\xi}v_{FG}.
\eea
The $v_{FG}=\sqrt{\013}v_f$ is that for the ideal
    non-interacting fermion gas with $v_f={k_f}/{m}$ being the Fermi speed.
A detailed information of the sound speed of the homogeneous system
    is very crucial for determining the frequencies of collective modes
    in the realistic trapped systems.

With $T=0$, the ratio of pressure to binding energy is still $\023 $
    of the non-interacting fermions gas in the unitary limit as pointed
    out in the literature\refr{Ho2004}.
The HvH theorem ensures the thermodynamic consistency dealt with
    the in-medium interaction potential in addressing the strongly
    interacting physics.
The same ratio is counterintuitive to some extent, which indicates
    that the universal thermodynamic properties with the infrared long
    range quantum fluctuation/correlation in the strongly interacting limit are
    extremely profound.

The numerical results for $T\neq 0$ can be readily given in the
    unitary limit regime with the explicit analytical expressions,
    from which one can see that the collective strongly correlating and
    rearrangement effects influence the chemical potential and pressure
    simultaneously.
However, the internal energy and Helmholtz free energy densities do
    not directly rely on the intriguing rearrangement effect even
    with a careful numerical check.

\section{Discussion and prospective}\label{sec5}

In addressing the unitary fermions physics,
     the essential task is how to incorporate the significant collective density and
    temperature fluctuation/correlation contributions into the
    thermodynamics in a reasonably controllable way.

The physical connotations are crystal clear in this analytical
attempt:
On one hand, the many-body surrounding environment
    modulates the bare two-body interaction,
    which is characterized by the in-medium renormalized
    interaction matrix $U_{eff}^*[n,T]$.
On the other hand, the collective correlation effects further shift
    the single particle energy spectrum, which is described by the
    corresponding single particle correlation potential $\mu _r[n,T]$.
The crucial energy-momentum conservation law is guaranteed by the
    \textit{simultaneous} consideration of these two collective
    correlation effects.
The fundamental thermodynamic consistency is the basic validity
    requirement of any scientific theoretical approaches,
    which can be easily neglected.

The generic infrared singularity of the bare two-body scattering
    amplitude at unitarity provides the scaling basis,
    which leads to the thermodynamic universal hypothesis. This scaling property
    implies that the thermodynamic expressions can be very
    simple in principle, i.e, the analytical formulae should be similar to those for the ideal non-interacting fermions gas.
In turn, this infrared singularity makes any potential theoretical
    approaches nontrivial.
Essentially, the unitary characteristic rules out any
    naive expansions or conventional perturbative loop diagram
    ring/ladder resummation calculations.

In this work, we have obtained the $T=0$ dimensionless universal
    coefficient $\xi=\049$ for the $D=3$ unitary fermion gas.
It is in the regime of the reasonable simulation/experimental values
    $0.42-0.46$.
However, as discussed recently in the
literature\refr{Krippa2007,Kohler2007}, it is yet far
    from the ending of looking for its realistic exact
    value.

The exact determination of this fundamental constant will influence
    the detailed knowledge about the collective excitation modes
    measured in the realistic ultra-cold unitary atomic fermions.
Consequently, more and more theoretical/experimental efforts are in
    fact most expected to make it certain precisely.
To understand the comprehensive
    thermodynamic properties exhibited by these novel
trapped strongly interacting quantum systems, there is still a
faraway route before us to constitute a
    consistent microscopic theory and corresponding
    analytical calculational framework.

\section{Conclusion}\label{sec6}

In summary, based on the Dyson-Scwhinger equations formalism, the
    ground state of the $D=3$ ultra-cold unitary fermion gas has
    been calibrated in detail in terms of the quantum Ising-like Bethe-Peierls contact interaction
     model. The universal factor is calculated to be
    $\xi=\049$.

In addressing the universal resonant thermodynamics, the medium
    dependent interaction through a
    dynamically twisted minus sign caused by an assumed mirror background makes
    it possible for us to incorporate the Landau pole contributions with the Fermi surface
    instability.
With the spontaneously generated rearrangements due to the single
    particle energy spectrum shift,
    the additional instantaneous correlation potential ensures the classical thermodynamic consistency.
Finally, we want to emphasize that there are not any
    infrared and/or ultraviolet cutoffs or regularization and/or expansion parameters
    obviously violating the significant unitarity, i.e., the bare
    $U_0$ drops out in the unitary limit.

\acknowledgments{The authors acknowledge the suggestive comments of
the referee(s). J.-S Chen thanks Lianyi He for help.
    Supported in part by the starting research fund of Central China
    Normal University and NSFC.}


\begin{thebibliography}{99}

\bibitem{Bertsch1999}
R.A. Bishop, \ijmpb{15}{{\it iii}}{2001}.

\bibitem{Baker2001}
G.A. Baker, \ijmpb{15}{1314}{2001}.

\bibitem{Heiselberg2000}
H. Heiselberg, \pra{63}{043606}{2001}.

\bibitem{Ho2004}
T.-L. Ho, \prl{92}{090402}{2004}.

\bibitem{Schwenk2005}
A. Schwenk and C.J. Pethick, \prl{95}{160401}{2005}.

\bibitem{Xiong2005}
H.-W. Xiong, S.-J. Liu, W.-P. Zhang, and M.-S. Zhan,
\prl{95}{120401}{2005}.

\bibitem{Horowitz2006}
C.J. Horowitz and  A. Schwenk, \plb{638}{153}{2006}.


\bibitem{physics/0303094}
J. Carlson,  S.Y. Chang,  V.R. Pandharipande, and K.E. Schmidt,
\prl{91}{050401}{2003}; S.Y. Chang, V.R. Pandharipande, J. Carlson,
and K.E. Schmidt, \pra{70}{043602}{2004}.

\bibitem{Castin2006}
Y. Castin, e-print arXiv:cond-mat/0612613.

\bibitem{Stringari2007}
S. Stringari, e-print arXiv:cond-mat/0702526.

\bibitem{Bloch2007}I. Bloch, J. Dalibard, and W. Zwerger, e-print
arXiv:0704.3011.

\bibitem{Pethick}
 C.J. Pethick and H. Smith, {\em Bose-Einstein Condensation in Dilute
Gases} (Cambridge University Press, Cambridge, England, 2002).

\bibitem{Steele}
J.V. Steele, e-print arXiv:nucl-th/0010066.


\bibitem{Astrakharchik2004}
G.E. Astrakharchik,  J. Boronat,  J. Casulleras, and  S. Giorgini,
\prl{93}{200404}{2004}.

\bibitem{Bulgac2005}
A. Bulgac, J.E. Drut, and P. Magierski, \prl{96}{090404}{2006};
e-print arXiv:cond-mat/0701786.

\bibitem{Hu2007}
H. Hu, X.-J. Liu, and P.D. Drummond, \epl{74}{574}{2006}; e-print
arXiv:cond-mat/0701744.


\bibitem{Partridge2006}
G.B. Partridge, W. Li, R.I. Kamar, Y.-A. Liao, and R.G. Hulet,
\sci{311}{503}{2006}.

\bibitem{Stewart2006}
J.T Stewart, J.P. Gaebler, C.A. Regal, and D.S. Jin, Phys. Rev.
Lett. 97, 220406 (2006).

\bibitem{Joseph2006}
J. Joseph, B. Clancy, L. Luo, J. Kinast, A. Turlapov, and J.E.
Thomas, \prl{98}{170401}{2007}.


\bibitem{chen2006}J.-S. Chen, \cpl{24}{1825}{2007} and \ctp{48}{99}{2007}.

\bibitem{chen2005}
J.-S. Chen, J.-R. Li, and M. Jin, \plb{608}{39}{2005}.

\bibitem{chen2005-1}
J.-S. Chen, e-print arXiv:nucl-th/0509038.

\bibitem{chen2005-2}
J.-S. Chen, J.-R. Li, and P.-F. Zhuang, \prc{67}{068202}{2003}.


\bibitem{walecka1974}
J.D.~Walecka, \ann{83}{491}{1974}.

\bibitem{Serot1986}
B.D.~Serot and J.D.~Walecka, \anp{16}{1}{1986};
    \ijmpe{6}{515}{1997}.

\bibitem{Hugenholtz}N.M Hugenholtz and L.van Hove, Physica (Amsterdam) {\bf 24}, 363
(1958).

\bibitem{Brown2002}G.E. Brown and M. Rho, \prep{363}{85}{2002}.

\bibitem{Baker1971}G.A. Baker, \rmp{43}{479}{1971}.

\bibitem{Rios2005}
A. Rios, A. Polls, A. Ramos, and I. Vida$\tilde{n}$a,
\prc{72}{024316}{2005}.

\bibitem{Nakada2006}
H. Nakada and T. Shinkai, e-print arXiv:nucl-th/0608012.

\bibitem{Kohn1965}
W. Kohn and L.J. Sham, \pra{140}{1133}{1965}.


\bibitem{Landau}L.D. Landau and E.M. Lifshitz, \textit{Statistical
Physics} (Pergamon Press, 1980).

\bibitem{Walecka}A.L Fetter and J.D Walecka, \textit{Quantum Theory of Many-Particle
Physics} (McGraw-Hill, New York, 1971).

\bibitem{chen2002}
J.-S. Chen, P.-F. Zhuang, and J.-R. Li, \prc{68}{045209}{2003};
\plb{585}{85}{2004}.

\bibitem{Krippa2007}B. Krippa, e-print arXiv:0704.3984.

\bibitem{Kohler2007}H.S. Kohler, e-print arXiv:0705.0944.






\end{thebibliography}
\end{document}